\documentclass[12pt,preprint]{aastex}

\shorttitle{The radio-loud NLQSO SDSS\,J172206.03+565451.6}
\shortauthors{S. Komossa et al.}

\begin{document}

\title{The radio loud narrow-line quasar SDSS\,J172206.03+565451.6}

\author{Stefanie Komossa, Wolfgang Voges, Hans-Martin Adorf}  
\affil{Max-Planck-Institut f\"ur extraterrestrische Physik,
Postfach 1312, 85741 Garching, Germany; skomossa@mpe.mpg.de}

\and

\author{Dawei Xu}
\affil{National Astronomical Observatories, Chinese Academy of Science, 
A20 Datun Road, Chaoyang District, 
Beijing 100012, China}

\and
\author{Smita Mathur} 
\affil{Department of Astronomy, The Ohio State University, 140 West 18th Avenue, Columbus, OH 43210, USA}

\and
\author{Scott F. Anderson}
\affil{Department of Astronomy, University of Washington, 
        Box 351580, Seattle, WA 98195 }

\begin{abstract}

We report identification of the radio loud narrow-line quasar 
SDSS\,J172206.03+565451.6
which we found in the course of a search for radio loud
narrow-line 
Active Galactic Nuclei (AGN).
SDSS\,J172206.03+565451.6 is only the $\sim$4th securely identified
radio loud narrow-line quasar and the second-most radio loudest
with a radio index $R_{1.4}\approx{100-700}$. 
Its black hole mass, $M_{\rm BH} \simeq (2-3) \times 10^{7}$
M$_{\odot}$ estimated from H$\beta$ line width 
and 5100\AA~luminosity, is unusually small given its radio loudness,  
and the combination of mass and 
radio index puts SDSS\,J172206.03+565451.6
in a scarcely populated region of $M_{\rm BH}$-$R$ diagrams.  
SDSS\,J172206.03+565451.6 is a classical Narrow-Line Seyfert\,1-type object
with FWHM$_{\rm H\beta} \simeq 1490$ km/s, 
an intensity ratio of [OIII]/H$\beta \simeq 0.7$ and FeII emission complexes
with FeII$\lambda$4570/H$\beta \simeq 0.7$. 
The ionization parameter of its narrow-line region, estimated
from the line ratio [OII]/[OIII], is similar to Seyferts and 
its high ratio of [NeV]/[NeIII] indicates a strong EUV to soft-X-ray excess.
We advertise the combined usage of [OII]/[OIII] and [NeV]/[NeIII]
diagrams as a useful diagnostic tool to estimate ionization
parameters and to constrain the EUV--soft-X-ray 
continuum shape relatively independent
from other parameters.  
\end{abstract}


\section{Introduction}

Narrow-line Seyfert 1 (NLS1) galaxies
were identified by Osterbrock \& Pogge (1985) as a
subgroup of Seyfert galaxies with small widths
of the broad Balmer lines (FWHM$_{\rm H\beta} <$ 2000 km/s).
Correlation analyses have shown that   
H$\beta$ line width is 
anti-correlated with the strength of FeII/H$\beta$ emission complexes
and correlated with the strength of [OIII]$\lambda$5007 (Boroson \& Green 1992). 
The study of NLS1s and correlations among their properties
is expected to hold
important clues on the  physical
processes in the central region of AGN (see Veron-Cetty \& Veron 2000
for a general AGN review).
There is increasing evidence that most NLS1s accrete close
to or at the Eddington limit   
(e.g., Boroson \& Green 1992, Pounds et al. 1995, 
Marziani et al. 2001,  
Simkin \& Roychowdhury 2003, Xu et al. 2003, Wang \& Netzer 2003, Grupe 2004,  
Collin \& Kawaguchi 2004)
and that they follow a different black hole mass -- velocity dispersion ($M_{\rm BH}-\sigma$)
relation than broad line Seyfert 1 galaxies 
(e.g., Mathur 2001, Grupe \& Mathur 2004, Botte et al. 2004, Mathur \& Grupe 2005).  

While the optical and X-ray properties of  
NLS1s were explored intensively in the last decade, 
relatively little is known about their radio properties (Ulvestad et al 1995,
Moran 2000, Greene et al. 2005). 
In particular, 
only a few radio loud NLS1s and narrow-line quasars (NLQSOs)   
have been identified so far{\footnote{there      
are preliminary reports that optical identifications
of NVSS radio sources produce a higher rate of radio loud NLS1s
than optical selection (Whalen et al. 2001)}}, 
namely three detections and one or two candidates.
PKS0558-504 (Remillard et al. 1986, Siebert et al. 1999),
RXJ0134-4258 (Grupe et al. 2000), and
SDSS\,J094857.3+002225 (Zhou et al. 2003) all match 
optical classification criteria of NLQSOs
and show radio indices exceeding $R=10$,
SDSS\,J094857.3+002225 even has $R > 1000$.
PKS 2004-447 (Oshlack et al. 2001)  
may be a non-typical NLS1 or possibly
a narrow-line radio galaxy (Zhou et al. 2003)
and is very radio loud.
RGB J0044+193 (Siebert et al. 1999)
is radio quiet most of the time (Maccarone et al. 2005), 
but may have been radio loud at
the epoch of the 87GB survey.

The reason for the scarcity of radio loud 
NLS1s is still unknown.
Finding new radio loud NLS1s is thus
of great importance for understanding the NLS1 phenomenon.
According to Boroson \& Green (1992), 
radio loudness in their quasar sample 
is closely linked to ``eigenvector 1'' in the sense that
radio sources tend toward {\em small} values of FeII equivalent width
and high [OIII] peak values (see also Marziani et al. 2001). 
The study of radio properties of NLS1s, in particular their radio loudness
and radio variability,  
also allows us to re-address questions related to the orientation of 
the inner region of NLS1s. It was originally suggested that NLS1s
are preferentially viewed face on (Osterbrock and Pogge 1985)
and some evidence for and against this scenario has been presented since then. 
The orientation issue is important not only for 
its own sake, but also enters black hole
mass and accretion rate estimates and consequently
affects black hole mass -- bulge mass
relations of NLS1s and their cosmological implications (e.g. see the discussion
in Collin \& Kawaguchi 2004). 
Finding more radio loud NLS1s may also shed new light onto the radio loud -- 
radio quiet dichotomy of AGN (e.g., Sulentic et al. 2003, McLure \& Jarvis 2004,
and references therein),
one of the key unsolved problems in AGN physics.

Here, we report detection of the radio loud NLQSO  SDSS\,J172206.03+565451.6,
found in the course of a search for radio loud narrow-line AGN (Komossa et al. 2005)  
making use of tools developed for the German Astrophysical Virtual
Observatory GAVO{\footnote{http://www.g-vo.org/portal/}}. SDSS\,J172206.03+565451.6
is exceptional in that it is only the fourth identified NLQSO,  
is the second-most radio loud, and  
shows other interesting properties including 
a high accretion rate and unusually low BH mass despite its radio loudness.  
In our terminology, we follow the classical distinction 
between narrow-line Seyfert 1 galaxy and narrow-line quasar
according to absolute visual magnitude when referring
to individual objects,
but we will collectively speak of NLS1 galaxies when referring to
class properties which then includes both, NLS1s and NLQSOs.
A cosmology with $H_{\rm 0}$=70 km/s/Mpc, $\Omega_{\rm M}$=0.3
and $\Omega_{\rm \Lambda}$=0.7 is adopted throughout, unless noted otherwise.  
 
In Section 2 we establish SDSS\,J172206.03+565451.6
as a radio loud object and present results from optical and X-ray spectroscopy,
followed by a discussion in Section 3.

\section{Radio, optical and X-ray observations of SDSS\,J172206.03+565451.6}

\subsection{Radio-loudness of SDSS\,J172206.03+565451.6}

SDSS\,J172206.03+565451.6 at redshift $z$=0.425  
was identified (Anderson et al. 2003, Williams et al. 2002)
in the course of the SLOAN Digital Sky Survey 
(SDSS; e.g., York et al. 2000) and 
was detected as X-ray source during the
{\sl ROSAT} All-Sky Survey (RASS, Voges et al. 1999).
It was classified as NLS1 
(Williams et al. 2002) but little else is known about this galaxy. 

SDSS\,J172206.03+565451.6 stuck out as exceptionally radio loud in a 
search for radio loud narrow-line AGN 
by cross-correlating the NLS1s in the 
``Catalogue of Quasars and AGN'' (Veron-Cetty \& Veron 2003)   
with different radio catalogues including the FIRST, NVSS, WENSS,
SUMSS and other surveys (for cross-correlation techniques and 
results on the full sample see Komossa et al. 2005). 
SDSS\,J172206.03+565451.6 is 
detected in the NVSS, FIRST and WENSS survey (Tab. 1). 
The radio positions agree within $<$1$^{\prime\prime}$
(WENSS: within 2.5$^{\prime\prime}$) with the optical position
of the galaxy.  
No indications for radio variability are seen, comparing
the FIRST and NVSS radio flux. 
Zhou et al. (2003) mentioned in passing the radio loudness of this galaxy
but did not discuss it further.

SDSS\,J172206.03+565451.6 shows a large scatter in blue magnitude
when comparing different epochs (Tab. 1), 
$\Delta$m$_{\rm B} \simeq 2.1$ mag, strongly indicating
that the variability is real. At faintest state, m$_{\rm B2}$=19.92 mag.
Using SDSS $g^{\prime}$ and $r^{\prime}$ magnitudes (Tab. 1),
we obtain m$_{\rm B_{SDSS}}$=18.46  with the color transformation
according to Smith et al. (2002). 

Following the conventions of  Kellermann et al. (1989),
we calculate the radio index $R$ as ratio of 
6 cm radio flux to optical flux at 4400\AA.
Under the assumption of similar spectral shapes in the optical
and radio band with spectral index $\alpha=-0.5$ (Kellermann et al. 1989),
$R=10$ approximately marks the ``border'' between radio quiet and radio loud objects
(which corresponds to a radio index at 1.4 GHz of $R_{1.4}$= 1.9$R$).
For SDSS\,J172206.03+565451.6 we obtain $R_{1.4}=$100-700, depending 
on choice of blue magnitude m$_{\rm B1}$ and m$_{\rm B2}$ 
from the USNO-B1 catalogue (Tab. 1), and after carrying out a Galactic extinction
correction of A$_{\rm B}$=0.11 mag.  
This puts SDSS\,J172206.03+565451.6 well above the
threshold commonly used to call an object radio loud. 

\subsection{Optical spectroscopy}

In order to confirm the optical spectral 
classification of SDSS\,J172206.03+565451.6, and
to measure its continuum and emission-line properties,
we analyzed the SDSS spectrum of this galaxy.
The Third Data Release (DR3, Abazajian et al. 2005) processed 
spectrum was used{\footnote{data were retrieved
at http://cas.sdss.org/dr3/en/tools/explore/obj.asp}}. 
This spectrum is flux- and wavelength-calibrated by the spectroscopic
pipeline in the course of the DR3 processing.
We further applied a Galactic extinction correction to the spectrum,
using E$_{B-V}$=0.026 mag (Schlegel, Finkbeiner \& Davis 1998)
and a standard R=3.1 extinction law.  
The IRAF package SPECFIT (Kriss 1994) was used for spectral analysis. 

The spectrum of SDSS\,J172206.03+565451.6
is AGN dominated. We fit a single power law
to the underlying continuum using the "continuum windows" known to be 
relatively free from strong emission lines at 3010-3040, 3240-3270, 3790-3810,
4200-4230, 5080-5100, 5600-5630, 5970-6000, and 6005-6035~$\AA$ (Forster et al. 2001,
Vanden Berk et al. 2001). The continuum is well represented 
by a single powerlaw with index $\alpha=-1.38$ (Fig. 1), where 
$f_{\lambda} \propto \lambda^{+\alpha}$, except at highest energies
around MgII,
where the conntiuum level is slightly overpredicted. 

The optical-UV spectrum shows emission from FeII complexes.
An FeII spectrum, using the technique and optical FeII template described
in Boroson \& Green (1992), was first fitted and subtracted (excluding the 
UV FeII multiplets around MgII$\lambda$2798, not included in the template).

More than 10 emission-lines are clearly present in the spectrum,
including MgII2798, [NeV]3426, [NeIII]3869, [OII]3727, [OIII]5007
and Balmer lines.
The FeII- and continuum-subtracted spectrum was used to measure
emission line parameters (and a more appropriate local continuum around MgII2798
was specified to measure that line). 
Emission lines were fit by single- 
or multi-component Gauss and/or Lorentz profiles 
and emission-line widths and emission-line ratios were measured.
Results are listed in Tab. 2. Line-widths were corrected for
the instrumental broadening. The SDSS pipeline provides the resolution 
at every pixel. The instrumental response of relevant 
emission-lines in the spectrum is in the range of FWHM$_{\rm inst}$ = 
133 -- 178~km/s. It was corrected for in all FWHMs reported below.   

Most weak lines are similarly well fit by either a Gaussian
or Lorentzian line profile. In Tab. 2, therefore generally 
only results from
the Gaussian fits are reported, plus results from direct
measurements of line widths and line fluxes 
(without any assumption on line profile
shape). 
The profiles of MgII, [OIII] and H$\beta$ appear more
complex. 
[OIII]$\lambda$5007 shows an asymmetric broad wing
frequently observed in NLS1 galaxies
(e.g., Xu et al. 2003). The profile is well matched by two
Gaussian components (Tab. 2) the peaks of which show a relative shift
of 144 km/s. 
A direct measurement of
the full width at the half of the peak flux
gives FWHM$_{\rm[OIII]}$ = 425 km/s.  
The profile of H$\beta$ is best fit by a 
model consisting of two Gaussians{\footnote{There is some ongoing
discussion as to which profiles are the best representation
of the complex AGN emission line shapes (e.g., Veron-Cetty et al. 2001,
Sulentic et al. 2002). 
In the present work,
we applied both, combinations of Gaussians and Lorentzians to
the line profiles. We find that ultimately the profile
of H$\beta$   
is best represented by two Gaussians. However, explicit choice
of line profile does not have any further consequences for our study,
since derived line ratios and FWHMs for 
the brighter emission lines agree within
typically $\sim$30\% of each other, independent of profile choice.
The classification of SDSS\,J172206.03+565451.6 
as a NLS1 is unaffected. }}.   
A direct measurement of the line width provides 
FWHM$_{\rm H\beta,d}$ = 1494km/s.     

In summary, SDSS\,J172206.03+565451.6 fulfills the classical 
optical criteria
for classification as a NLQSO.
In particular, FWHM$_{\rm H\beta} \simeq$ 1494 km/s while
FWHM$_{\rm [OIII]} \simeq$ 425 km/s,
and $I_{\rm [OIII]\lambda5007}$/$I_{\rm H\beta_{total}} = 0.7.${\footnote{The "direct"
measurements ("d") of Tab. 2 are quoted here, since they are the most model-independent.}}
The ratio of FeII$\lambda$4570 to H$\beta_{\rm total}$ amounts to 0.7, 
where the FeII blend is measured between 4434\AA~ and 4684\AA.

The radio loudness of SDSS\,J172206.03+565451.6
can be re-evaluated by using the optical 
flux at 4400\AA ~restframe obtained from the SDSS spectrum
and the 5 GHz restframe radio flux
predicted from the actually observed radio spectral
index 
$\alpha_{\rm r_{1.4-0.33}}=-0.69$ ($f_{\rm \nu} \propto \nu^{\alpha}$). 
$\alpha_{\rm r}$ was calculated from the radio observations
at 1.4 (FIRST) and 0.33 GHz (Tab. 1) assuming
that no spectral break occurs in between.
This then gives $R_{\rm 5GHz}$=360 and re-confirms the radio loudness of
SDSS\,J172206.03+565451.6.

\subsection{X-ray spectroscopy and variability}    

SDSS\,J172206.03+565451.6 was detected in X-rays during  
the RASS. 
We analyzed these data in order to 
check for variability and
to measure the spectral shape (pointed observations with {\sl ROSAT}
or any other X-ray mission are not available). 
Standard procedures of data reduction were followed. 
Source photons were extracted from a circular region centered
on the source position. The background was determined in two source-free areas
along the scan direction of the X-ray telescope.
Data were corrected for vignetting.  

We find that the spectrum can be well fit by a single powerlaw,
but powerlaw index and amount of absorption are not well
constrained. If absorption is treated as free parameter, we obtain
$N_{\rm H} \simeq 0.45 \times 10^{21}$ cm$^{-2}$ and $\Gamma_{\rm{x}} \simeq -3.0$
($\chi{^{2}}_{\rm red}$=0.4), while for 
$N_{\rm H}$=$N_{\rm Gal}$=$0.16 \times 10^{21}$ cm$^{-2}$ we 
find $\Gamma_{\rm{x}} \simeq -2.1$ ($\chi{^{2}}_{\rm red}$=0.5).  
The absorption-corrected (0.1--10)\,keV flux obtained for
these two models is uncertain by a factor $\sim$2 and converts
into an X-ray luminosity of $L_{\rm (0.1-10)keV} \simeq (1-2) \times 10^{45}$ erg/s.   
More complex models than simple powerlaws 
are not warranted, given the faintness of the source
($\sim$115 source photons detected).   

We also inspected the X-ray lightcurve of SDSS\,J172206.03+565451.6
and find that its X-ray emission is constant within the errors
throughout the RASS observation. 

\section{Discussion}

\subsection{Radio properties of SDSS\,J172206.03+565451.6 and 
          comparison with other radio loud NLQSOs} 

We have shown that SDSS\,J172206.03+565451.6 fulfills all 
criteria for classification as a NLQSO and is
one of the radio loudest NLQSOs known.
This does not only hold for radio index, but also
for radio power.
We obtain $P_{\rm 1.4} \simeq 3 \times 10^{25}$ W/Hz, significantly
within the radio loud regime ($P_{\rm 4.85} > 10^{24}$ W/Hz,
Joly et al. 1991).
In particular, this is also orders of magnitude  
above the most radio-luminous starbursts studied by 
Smith et al. (1998) which have $\log P_{\rm 4.85,SB} \simeq 22.3-23.4$ W/Hz. 
This excludes the possibility that the radio emission of 
SDSS\,J172206.03+565451.6 is starburst-dominated.  
Also, we measured the ionization
parameter for SDSS\,J172206.03+565451.6 from [OII]/[OIII] which is in a regime
typical of Seyfert galaxies (next Section).
 
The radio spectral shapes of the few known radio loud NLQSOs  
appear to scatter widely,  with
$\alpha_{\rm 2.7,4.85 GHz}=0.6$ 
(SDSS\,J094857.3+002225, Zhou et al. 2003), 
$\alpha_{\rm 4.85,8.4 GHz}=-1.4$ (RXJ0134-4258, Grupe et al. 2000),
and $\alpha_{\rm 0.33,1.4 GHz}=-0.7$ (SDSS\,J172206.03+565451.6), while 
the NLS1-like galaxy PKS 2004-447 shows 
$\alpha_{\rm ATCA}=-0.6$ (Oshlack et al. 2001)
where $f_{\rm \nu} \propto \nu^{+\alpha}$. 
It has to be kept in mind, though, that they were measured
at different frequencies and are generally based on non-simultaneous
observations. 

While PKS0588-504 is highly variable in the X-ray band on various time scales
(e.g., Gliozzi et al. 2001), and RXJ0134-4258
dramatically changed its spectral shape
(Grupe et al. 2000, Komossa \& Meerschweinchen 2000),
little can be said on the longer-term X-ray variability properties
of SDSS\,J172206.03+565451.6 based on the short RASS observation. 
Dedicated X-ray observations of this galaxy will
be useful to re-address this issue.

\subsection{Emission-line diagnostics}   

We attempted to measure
extinction by using the Balmer lines H$\beta$, H$\gamma$ and H$\delta$.
However, at location of H$\delta$ the red and blue part of
the SDSS spectrum were merged, and a bump underlying and redward
of H$\delta$
implies  
its measurement is highly uncertain.
Similarly, H$\gamma$ comes with some uncertainty, because
the line is partially blended with [OIII]4363.
The observed value, H$\beta$/H$\gamma \simeq$ 0.28..0.46, is
uncertain by almost a factor of two, and within this
uncertainty is consistent with the Case-B recombination value, 0.47.
The absence of strong reddening of the spectrum is 
also hinted by the blue continuum exhibited
by SDSS\,J172206.03+565451.6.

The emission-line ratio [OII]$\lambda$3727/[OIII]$\lambda$5007 is
a good indicator of the ionization parameter $U$ (Penston et al. 1990), as long as the
emission-line region in question does not contain strong density inhomogeneities
(Komossa \& Schulz 1997). The ratio observed in SDSS\,J172206.03+565451.6,
log [OII]/[OIII] $\simeq -0.6$ (Tab. 2), 
implies $\log U \simeq -2.3...-2.6$
which is based on re-calculation of Fig. 7 of Komossa \& Schulz (1997)
but using typical NLS1 ionizing continua
and a NLR cloud density in the range $\log n_{\rm H} = 2-3$. 
The NLS1 continua are composed of the mean spectral energy distribution
(SED) of the NLS1 NGC\,4051 outside the EUV--X-ray regime plus a systematically varying
additional EUV excess and varying steepness of the X-ray continuum spectrum
(see caption of Fig. 2 for details). 
Photoionization calculations were carried out using the code
{\em Cloudy} (Ferland et al. 1998) with assumption as in 
Komossa \& Schulz (1997; in brief: ionization-bounded clouds of constant density
and solar metallicity illuminated by a central point-like continuum source).
Fig. 2a demonstrates that 
for NLS1-like continua, [OII]/[OIII] is still a good diagnostic of
ionization parameter, relatively independent 
of continuum shape{\footnote{deviations
arise for extreme soft excess continua and high ratios of [NeV]/[NeIII]
for which the method should be re-evaluated}}. 
The ionization parameter estimated for SDSS\,J172206.03+565451.6
is comparable to other Seyferts and in particular
is  similar to the NLS1 galaxy NGC\,4051
($\log U \simeq -2.2...-2.5$ applying the same method; Komossa \& Fink 1997).

Examining other line ratios for their diagnostic value, 
we find that once the ionization parameter is known,
the ratio of [NeV]/[NeIII] is a very good diagnostic of the
EUV-soft X-ray continuum, relatively independent of 
density (Fig. 2b).  
SDSS\,J172206.03+565451.6 is unusual in its relatively strong NeV
emission which does indicate the presence of a high-energy
continuum excess, i.e., a spectral energy distribution
(SED) similar to continuum `c3'-`c4' (see Fig. 2).

Essentially all of the 128 NLQSOs listed in the 
``Catalogue of Quasars and AGN''
compiled by Veron-Cetty \& Veron (2003) have redshifts in a range
where both [OII]$\lambda$3727 and [NeV]$\lambda$3426 are
easily measurable by optical spectroscopy.   
The combination of the [OII]/[OIII] and [NeV]/[NeIII] diagrams is
thus a very useful tool in estimating ionization parameters
of NLS1s and constraining their EUV continua. These line-ratios
should be a beneficial addition in sampling the ``Eigenvector''  
parameter space of radio loud and radio quiet NLS1s,
complementing line ratios such as 
FeII/H$\beta$ and 
line profiles shapes already in use 
(e.g. Sulentic et al. 2003, Bachev et al. 2004).

\subsection{Black hole mass and accretion rate}

Using the H$\beta$ line width and X-ray luminosity of 
SDSS\,J172206.03+565451.6, we obtain
an estimate of black hole mass and accretion rate  
in terms of the Eddington rate.  

Assuming that the broad line region (BLR) clouds are
virialized (e.g., Wandel et al. 1999), the black hole mass can be estimated as
$M_{\rm BH} = G^{-1}\,R_{\rm BLR}\,v^{2}$.  
The BLR radius, $R_{\rm BLR}$,
can be determined from the optical luminosity at 5100\AA~ 
according to equation (6) of Kaspi et al. (2000), 
$R_{\rm BLR} = 32.9[{{\lambda\,L_{\lambda}{\rm(5100\AA)}}\over{10^{44}{\rm erg/s}}}]^{0.7}$ ld
(derived for a cosmology with $H_{0}=75$ km/s/Mpc and $q_{0}=0.5$). 
The velocity $v$ of the BLR clouds is usually estimated
from the FWHM as $v = f\,{\rm FWHM}$ where $f={\sqrt{3}\over{2}}$ for
an isotropic cloud distribution. 
With $L_{\lambda}{\rm(5100\AA)_{\rm{H_0=75,q_0=0.5}}}=4.3 \times 10^{40}$ erg/s/\AA~ 
derived from the SDSS
spectrum,   
and FWHM$_{\rm H\beta,d}$=1494 km/s,
we obtain a black hole mass of $M_{\rm BH} \simeq 1.9 \times 10^{7}$ M$_{\odot}$
for SDSS\,J172206.03+565451.6. More appropriately using the FWHM of the broad
component from the two-component fit to H$\beta$, 
we get $M_{\rm BH} \simeq 3.3 \times 10^{7}M_{\odot}$.{\footnote{For comparison,
Bian and Zhao (2004), who included SDSS\,J172206.03+565451.6 in a sample 
of galaxies for which they estimated BH masses, 
obtained $M_{\rm BH} \simeq 2.3 \times 10^{7}$ M$_{\odot}$ based on a 
single-component direct measurement of FWHM$_{\rm H\beta}$.}}    

SDSS\,J172206.03+565451.6 falls right into an unpopulated 
regime of the ``Laor diagram'' (Fig. 2 of Laor 2000)
which plots the relation between radio loudness and black hole
mass.
While larger samples filled up some originally
empty areas in the Laor diagram (e.g., Lacy et al. 2001,
Oshlack et al. 2001, 2002, Woo and Urry 2002, McLure \& Jarvis 2004),
SDSS\,J172206.03+565451.6 is still located in a
region relatively sparsely populated.
This either implies that SDSS\,J172206.03+565451.6 has an exceptionally  
low black hole mass given its radio properties, or that beaming
is important in explaining its radio emission with the consequence
that this NLQSO is seen face-on, and is thus an important
cornerstone object for future studies of inclination
effects in NLS1 galaxies (see Komossa et al. 2005 for further discussion).

With knowledge of BH mass and X-ray luminosity it is possible to make a rough
estimate of the accretion luminosity relative to the Eddington
value.
According to Maccarone et al. (2003, and references therein),
Galactic X-ray binaries in soft/high-state and AGN show {\em suppressed}
radio emission for relatively high accretion rates, in
the range of 0.01-0.1 times the Eddington rate. 
With $M_{\rm BH} \simeq (1.9-3.3) \times 10^{7}$ M$_{\odot}$ and 
$L_{\rm x,(0.1-10)keV} \simeq (1-2) \times 10^{45}$ erg/s, 
we obtain $L$/$L_{\rm Edd}$=0.2-0.8 
{\em without} any bolometric correction which 
should typically be an additional factor 5--10
(Tab. 14 and 15 of Elvis et al. 1994).  
This estimate demonstrates that SDSS\,J172206.03+565451.6 
accretes at a high rate and, in particular, that it exceeds the accretion
rate at which radio emission
is expected to be suppressed.

\section{Summary}

We have presented a study of the radio loud
NLQSO SDSS\,J172206.03+565451.6. One of the radio loudest NLQSOs known, 
its combination of black hole mass and radio index
put it into a scarcely populated region in $M_{\rm BH}$-$R$
diagrams in that its black hole mass is unusually low
given its radio loudness. Its Eddington ratio $L/L_{\rm Edd}$ is
close to unity.    
Future searches for and study of similar objects may hold
important clues to the nature of radio loud NLS1s in
particular and NLS1 models in general,
and the mechanism which causes the radio loud -- radio quiet
bimodality in AGN.

\acknowledgments
 We thank Gary Ferland for providing {\em Cloudy}.
 GAVO is funded by the {\sl Bundesministerium f\"ur Bildung und Forschung} (BMBF)  
 under contract no. 05\,AE2EE1/4. DX acknowledges the support of the Chinese 
 National Science Foundation (NSF) 
 under grant NSFC-10503005.    
 This research made use of the SDSS and Vizier archives, the NED service,
 and the Guide Star Catalogue-II.

\clearpage

\clearpage 

\begin{figure}
\plottwo{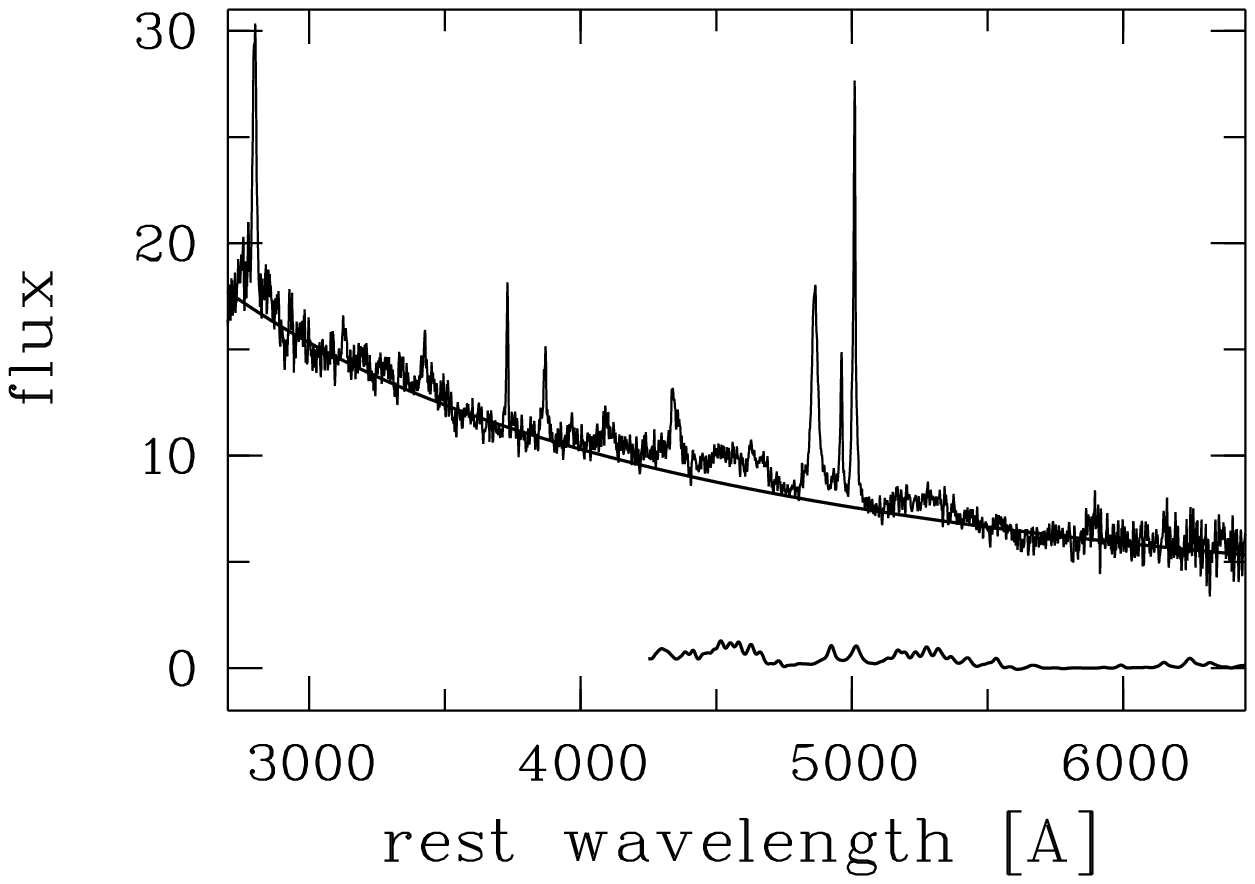}{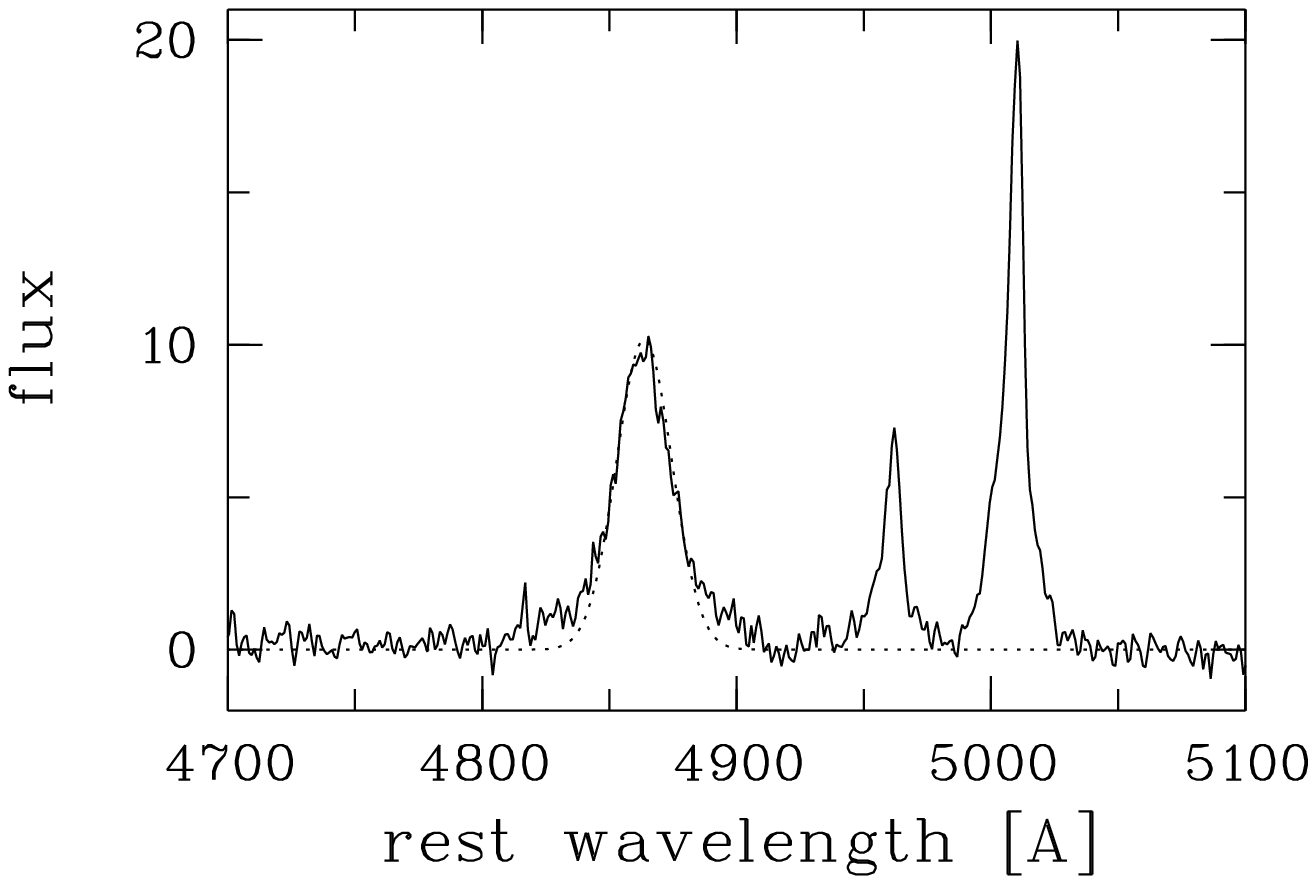}
\caption{Left: Optical spectrum of SDSS\,J172206.03+565451.6 
(flux $f_{\lambda}$ in arbitrary units)
and best-fit 
power-law continuum. The brightest lines are marked.
The spectrum is shifted to the rest frame
wavelength and smoothed with a boxcar of 3 pixels for
clarity.
The lower curve shows the best-fit FeII emission contribution.   
Right: Zoom onto the H$\beta$-[OIII] complex. The dashed line shows a 
single-Gaussian representation of H$\beta$.   
}
\end{figure}

\clearpage

\begin{figure}
\plottwo{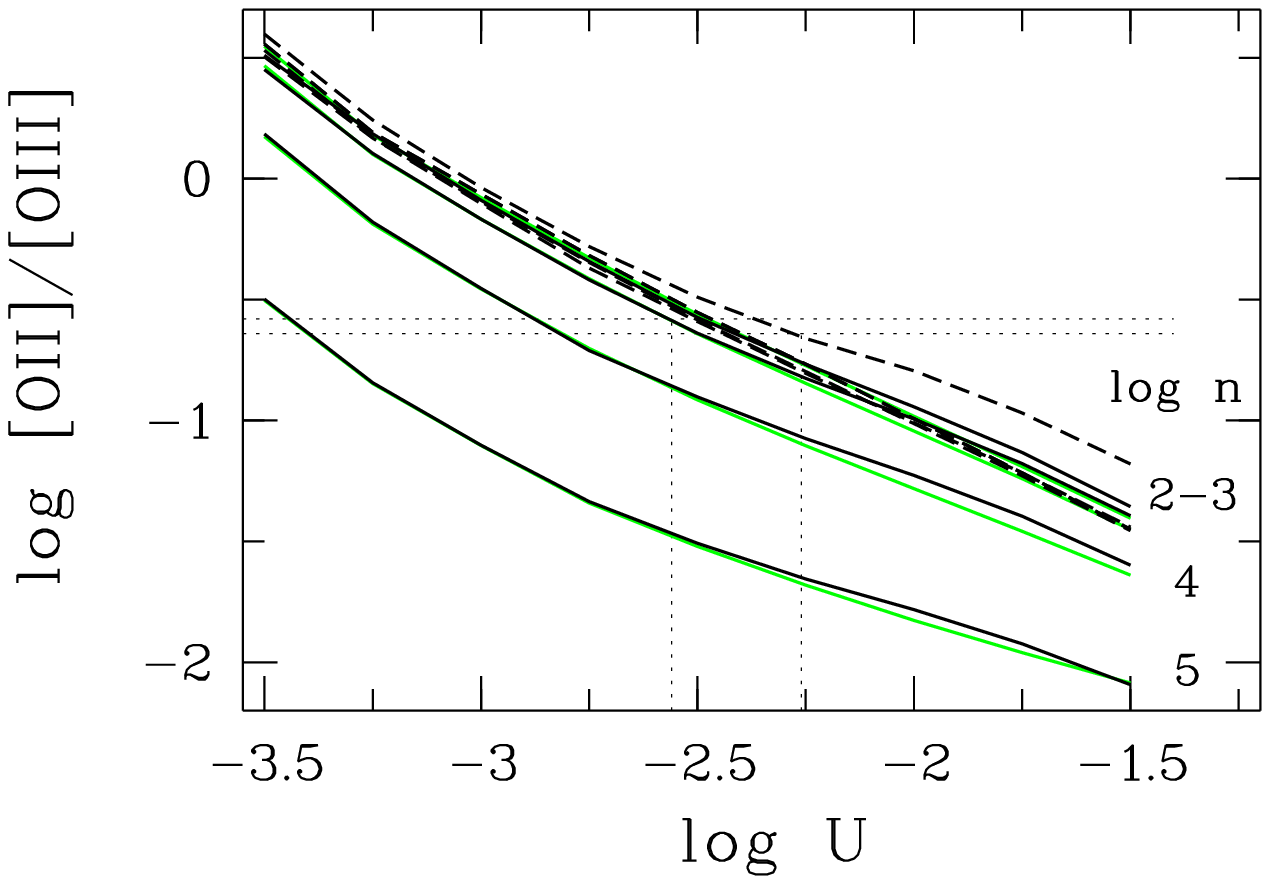}{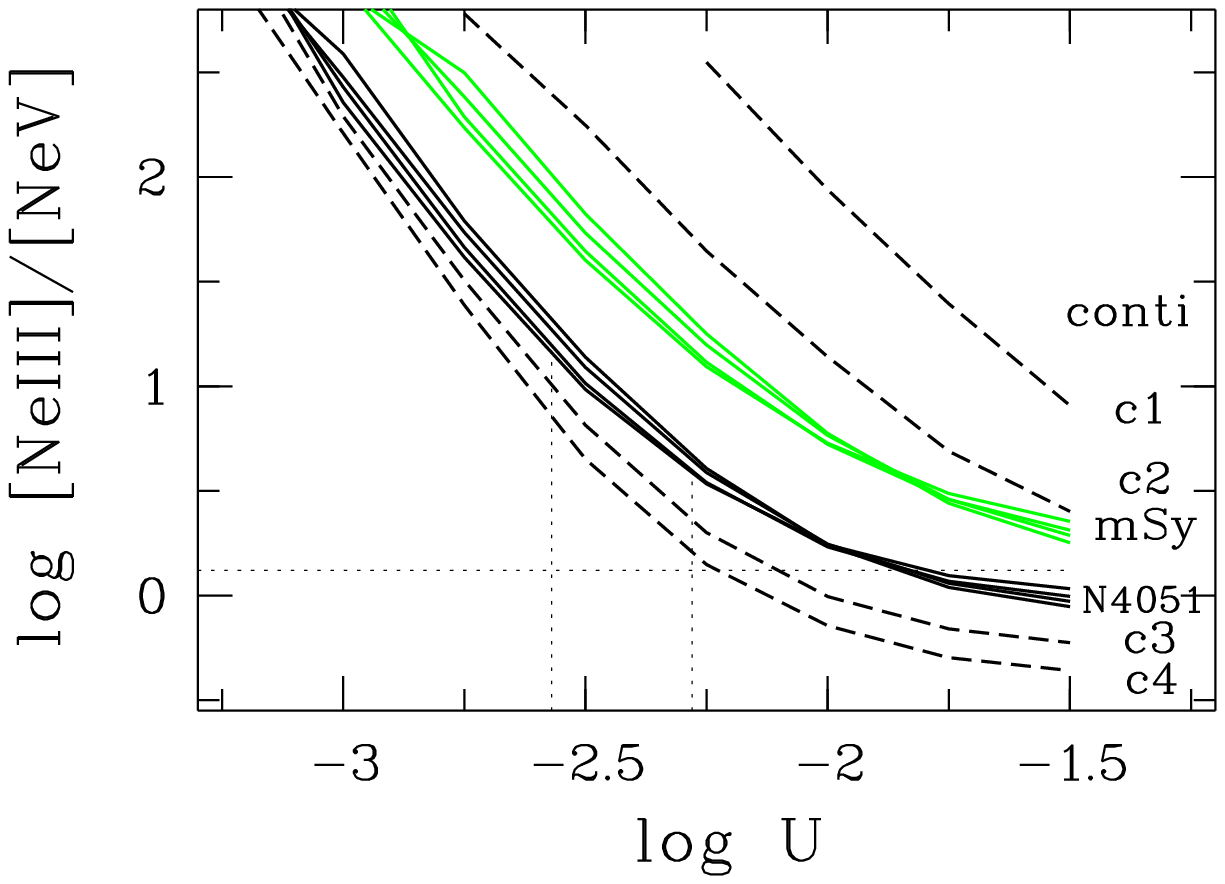} 
\caption{Left: The dependence of the emission-line ratio [OII]3727/[OIII]5007 
on the ionization parameter $U$ 
for different continuum shapes and cloud gas densities.  
The density varies from top to bottom
between $\log n$ = 2-5, as marked. 
The different continuum shapes lead to a very similar
variation in [OII]/[OIII] and we do not label them individually
(the black solid line corresponds to the multi-wavelength 
spectral energy distribution (SED) 
of the NLS1 galaxy NGC\,4051 (Komossa \& Fink 1997) while the dashed line was calculated for 
NGC\,4051-like continua with varying strength of an EUV--soft-X-ray excess,
and the grey line was calculated for a mean Seyfert continuum). Models with
$\log n$ = 4,5 were not re-calculated for all continuum shapes. 
The dotted line marks the measured [OII]/[OIII] ratio with 
its uncertainty.  \newline 
Right: The dependence of the emission line ratio [NeV]3426/[NeIII]3869 
on $U$ for different continuum shapes and gas densities.    
Line style coding is the same as for the [OII]/[OIII] diagram.
In addition, the continuum shape is indicated at the 
right-hand side of the figure. 
Models c1 to c4 correspond to NGC\,4051-like SEDs except that the 
flux at $\log \nu = 16.383$ Hz, connected by powerlaws
to the flux at the Lyman limit and to the flux at $\log \nu = 19.383$ Hz,
was systematically increased -- by factors of 0.04  
(= EUV deficit rather than excess), 0.1, 3 and 7 for c1-c4, respectively -- 
in order to 
describe soft excesses of varying strength.
This then also changed the powerlaw 
indices $\alpha_{\rm EUV}$ and $\alpha_{\rm x}$
while the
remainder of the SED was kept constant.
Models involving the SED of NGC\,4051 itself or a mean Seyfert `mSy' SED
were re-calculated for densities ranging between $\log n$ = 2-4.
Individual densities are not marked in the plot
since it can be seen that the density dependence is very weak.  
}   
\end{figure}

\clearpage

\begin{table}
\rotate
\begin{center}
\caption{Properties of SDSS\,J172206.03+565451.6. 
Flux/magnitude estimates are based on
the following catalogues:  
USNO (Monet et al. 2003), GSC 2.2, SDSS (Anderson et al. 2003),
FIRST (Becker et al. 2003), 
NVSS (Condon et al. 1998), and WENSS (de Bruyn et al. 1998). 
}
\begin{tabular}{lll}
\tableline
\tableline
waveband & flux/brightness & catalogue and comments  \\
\tableline
m$_{\rm B_1}$  & 17.8 mag & USNO-B1.0 \\
m$_{\rm B_2}$  & 19.9 mag & USNO-B1.0 \\
m$_{\rm B_{\rm J}}$ & 18.66 mag & GSC 2.2 \\
u$^{\prime}$  & 18.65 mag & SLOAN \\
g$^{\prime}$ & 18.35 mag & SLOAN \\
r$^{\prime}$ & 18.26 mag & SLOAN \\
i$^{\prime}$ & 18.07 mag & SLOAN \\
z$^{\prime}$ & 17.93 mag & SLOAN \\
1.4 GHz & 39.8 mJy & FIRST \\
1.4 GHz & 42.6 mJy & NVSS \\
0.33 GHz & 108 mJy & WENSS \\
X-rays, (0.1--10) keV   &  1.5..3.6\,10$^{-12}$ $^{*}$    & this paper \\  
\tableline
\end{tabular}
\end{center}
$^{*}$absorption-corrected flux in erg\,cm$^{-2}$\,s$^{-1}$\,Hz$^{-1}$  (see Sect. 2.3 on models). 
\end{table}

\clearpage

\begin{table}
\rotate
\caption{Optical-UV emission lines of SDSS\,J172206.03+565451.6. 
The second column, labeled ``m'', 
refers to the method the line width and flux was measured (g = Gaussian profile fit, 
l = Lorentzian, d = direct integration over the observed profile, without any model
assumption, g+g = two component Gaussian of which narrow and broad component
are reported in separate rows). 
}
\begin{tabular}{lcccl}
\tableline
\tableline
line identification & m & line ratio & FWHM$_{\rm c}$ & comment  \\
                    &   &         & [km/s] &  \\
\tableline
~[OIII]$\lambda$5007 & g & 0.57 & 487  & \\
                     & d & 0.75 & 425 & \\
~~~[OIII]$_{\rm narrow}$ & g+g & 0.22 & 262 & \\   
~~~[OIII]$_{\rm broad}$  & g+g & 0.54 & 1044 & \\
~H$\beta$                   & g & 0.94 & 1577 & underpredicts broad wings (Fig. 1)\\
                           & l & 1.31 & 1476 & overpredicts flux in the wings \\ 
                           & d & 1.11 & 1494 & \\
~H$\beta$$_{\rm total}$ & g+g & 1.0! & &   \\
~~~H$\beta$$_{\rm narrow}$ & g+g & 0.05 & 425! & FWHM$_{\rm H\beta_{\rm narrow}}$ fixed to FWHM$_{\rm [OIII]_{\rm d}}$ \\
~~~H$\beta$$_{\rm broad}$  & g+g & 0.95 & 1980 &  \\
~FeII~4570           &   & 0.73   &     &    \\
~[NeIII]$\lambda$3869   & g & 0.09 & 429 & \\
                         & d & 0.10: &  415 & \\ 
~[OII]$\lambda$3727     & g & 0.15 & 418 & \\
                        & d & 0.17 &  495 & \\ 
~[NeV]$\lambda$3426     & g & 0.08 & 599 &  \\
                        & g & 0.07 & 429! & FWHM$_{\rm [NeV]}$ fixed to FWHM$_{\rm [NeIII]_g}$ \\ 
~MgII$\lambda$2796+2803 & g & 0.80 & 1684 &   \\
                        & d & 0.87 & 1433 &   \\
\tableline
\end{tabular}
\end{table}


\end{document}